# Evolution and Detection of Polymorphic and Metamorphic Malwares: A Survey

Ashu Sharma
Research Scholar
Department of C.S. & I.S.
BITS Pilani, K K Birla Goa Campus, Zuarinagar,
Goa - 403726, India.

ashu.abviiitm@gmail.com

S. K. Sahay.
Assistant Professor
Department of C.S. & I.S.
BITS Pilani, K K Birla Goa Campus, Zuarinagar,
Goa - 403726, India.
ssahay@goa.bits-pilani.ac.in

## ABSTRACT
Malwares are big threat to digital world and evolving with high complexity. It can penetrate networks, steal confidential information from computers, bring down servers and can cripple infrastructures etc. To combat the threat/attacks from the malwares, anti- malwares have been developed. The existing anti-malwares are mostly based on the assumption that the malware structure does not changes appreciably. But the recent advancement in second generation malwares can create variants and hence posed a challenge to anti-malwares developers. To combat the threat/attacks from the second generation malwares with low false alarm we present our survey on malwares and its detection techniques.

## General Terms
Information security and Malware analysis.

## Keywords
Malwares, Antimalware, Polymorphic, Metamorphic.

## 1. INTRODUCTION
A malware is a malicious software/program/code which enters system without user authorization and takes undesirable actions. The term is too often used interchangeably with virus, even though the two are not the same. Malware is actually a condensed, conjoined term used to refer viruses, worms, trojan horses, spyware, adware, rootkits, botnets etc. In today's computing world malwares are a big threat and are continuously growing with high complexity. The reason behind the increase in threat from malware is the wide spread use of World Wide Web. An estimate shows that the web based attack increased 36% with over 4,500 new attacks each day, annoying/disrupting the victim in terms of confidentiality, integrity, availability of the victim's data etc. [1].

In 2011, Symantec Internet Security reported that ∼ 403 million new variants of malware were created, a 41% increase from 2010 [1]. State sponsored highly skilled hackers are developing customized malwares to disrupt industries and for military espionage [2]. Such attacks can alter the operation of industrial systems or disrupt power plants etc. [3] The intrusion into Google's systems demonstrates how well-organized attacks are designed to maintain long-term access of an organizations network [4].

Success of windows based malware has inspired attackers to develop cross-platform variants to maximize the damages. Therefore Linux OS are no more immune to the malware attacks. Over the years Linux features are more or less same, hence some rootkits that have been used decade back are still being used for the attacks, e.g. the Adore root kit, trojanized system binaries, SSH servers etc. [5].

Personal Digital Assistants (PDAs), in particular mobile phones, which are integral part of our lifestyle, are also vulnerable to malware attacks. F-secure, documented a recent increase in malware attacks against mobile devices based on Android and Apple iOS [6]. The McAfee threat report reveals a significant increase in mobile malwares from 2004 to 2012, and claimed more than 8,000 mobile malwares are collected in their databases [7]. According to Symantec Internet Security, out of 5,291 new vulnerabilities that have been discovered in 2012, 415 of them are on mobile operating systems [1].

Since the first virus created in 1970 [8], there is a strong contest between the attackers and the defenders. To defend the malware attacks, anti-malware groups are developing new techniques. On the other hand, malware developers are adopting new tactics/methods to avoid the malwares detectors. Initially the tools and techniques of malware analysis were in the domain of anti-malware vendors. However, the use of malware for espionage, sophisticated cyber-attacks and other crimes has motivated academicians and digital investigators to develop advanced methods to combat the threats/attacks from it.

There are many malware detection systems viz. Signature based detection, code emulation, heuristic code analysis and machine learning. Some malwares are easy to detect and can be removed from the system by commonly used signature based antivirus software. But the signature based technique can't detect new or previously unknown variant of malwares. This method of detection worked well until the malware group started developing the polymorphic and metamorphic malwares. Knowing the limitation of signature based detection technique, malware developers are creating variations in malwares by employing a variety of code obfuscation methods viz. reordering instructions, renaming registers, substituting sets of equivalent instructions and inserting junk snippets. Rest of the paper is organized as follows: Section 2 describes the types of Malwares. Section 3 discusses the detection techniques of malwares. Finally in section 4 we discuss the future direction of anti-malwares development.

## 2. TYPES OF MALWARES

Malwares are basically classified as first generation and second generation. In first generation, structure of the malwares does not change. But in second generation, the internal structure of malwares change in every variant while the actions are maintained same. On the basis of how variances are created in malware, second generation malwares are further classified as Encrypted, Oligomorphic, Polymorphic and Metamorphic Malwares.

### 2.1 Encrypted Malwares

Encryption was the first concealment techniques used for creating the 2nd generation malwares [9]. It consists of two parts; the encrypted body and a decryption code [10]. Usually the body is XORed with a key to make it difficult to detect. For each infection, encrypted malware makes the body unique by using different key to hide the signature. However, the decryption routine remain same, hence it can be detected by analyzing the decryptor. In this method when the malware code executes; first the decryption part is executed to decrypt the body of the malware and then the code is executed for the action. The first encrypted malware was CASCADE [11]. Later on using the CASCADE technique Win95/Mad and Win95/Zombie were created. The main motivation to use the encryption malware is to avoid static code analysis, delay the process of inspection, prevent tampering and avoid detection [12].

### 2.2 Oligomorphic Malware

The short comings of the encrypted malware led to the development of different concealment techniques. In Oligomorphic malwares decryptors are mutated from one variant to other. Initially this type of malware was capable of changing the decryptor slightly [9]. The simple method to create Oligomorphic malwares is to provide a set of different decryptors rather than one. At most this malware can generate few hundred different decryptors, e.g. Win95/Memorial had the ability to build 96 different decryptor patterns [13]. For its detection, signature based techniques can be applied by making the signature of all the decryptors. However, in general to detect Oligomorphic malwares, signature based techniques are not a good approach [10].

### 2.3 Polymorphic Malwares

In Polymorphic malwares, millions of decryptors can be generated by changing instructions in the next variant of the malware to avoid signature based detection [12]. It also consists of two parts; the first part is the code decryptor to decrypt the second part (body). During the execution of malware, mutation engine creates a new decryptor which is joined with the encrypted malware body to construct a new variant of malware [14]. Polymorphic malwares are created by using the obfuscation techniques (dead-code insertion, register reassignment, subroutine reordering, instruction substitution, code transposition/integration etc.) [9]. The first known polymorphic malware was 1260, written by Mark Washburn in 1990 [10]. Although, a large number of variants of decryptors can be created, but still signature scanning technique can be used to detect the malwares by identifying the original program with emulation technique [9].

### 2.4 Metamorphic Malwares

Metamorphic malwares are body-polymorphic [10], i.e. Instead of generating new decryptor, a new instance (body) is created without changing its actions. Similar to polymorphic malware, obfuscation techniques can be used to create new instances. It is believed that in future it will harm both computers and PDAs in large scale as it is almost impossible to detect by signature based techniques. Creating a true metamorphic malware without arbitrarily increasing the size is a challenging task. It has been shown that there are only few malwares exhibit true metamorphic behavior [15], e.g. Phalcon/Skism Mass-Produced Code Generator, Second Generation virus generator, Mass Code Generator and Virus Creation Lab for Win32 were claimed to be metamorphic but were not. The first metamorphic virus was created in 1998 called as Win95/Regswap [16]. In 2000, Win32/Ghost virus was created with 3628800 different variants [16]. One of the strongest metamorphic malware W32/NGVCK was created in 2001 with the help of Next Generation Virus Creation Kit (NGVCK).

## 3. DETECTION TECHNIQUES

To combat the threat/attacks from the malwares, softwares (anti-malware) are developed, which are mostly based on the assumption that the malware structure does not change appreciably. But the variant of 2nd generation malwares are very much different to each other, hence threat/attacks from such malwares to Computers and PDAs are increasing day by day. Therefore, there is a need that both academia and anti-malwares developers should continually work to prevent damage from the evolution of malwares. This section discus the various techniques used for the detection of malwares.

### 3.1 Signature based detection

Signature detection is the simplest and an effective way of detecting known malwares [17]. Once the malware is identified, unique sequences of bytes are extracted from it, which represents the sig- nature of the malware. This signatures are selected long enough to characterize a specific malware with respect to any other benign program, e.g. Worm/klez.E and Worm/MyParty.Ao signatures are 33be732d4000bd08104000e89eeaffff80bd08104000be7d2d40 00e849eaffff6a00e83500000064756d6d792e65786500653a5c 77696e646f77735c53795374656d33325c644c6c63616368655 c6464642e65786500ff254c404000ff25544040 and aa328cf2 4554d90b307c407eca9a4cf02a4d5a90000332c8b26904ffffb8 40f97f370080040e1fba0e00b409cd21b8014c001f027c546869 73c363616e042568d54562e2c876b0ffbf0420444f53 respectively [18]. This techniques scans the file in the sys-tem to find the defined malware signature, if found an alert of the presence of malware is sent, e.g. Aho-Corasick algorithm scan for the exact matching, hence a slight mismatch will escape detection [19]. Veldamna and Wu-Manber proposed the use of wildcard for detecting slight variance in the malwares. Some metamorphic malwares could be detected using the wildcard method, e.g. W32/Regswap [12].

It is easy to use, however requirement of scanning becomes costly as the database of malware signature is increasing very fast [20]. Also, it's a completely reactive technique, therefore unable to combat threats/attack from the new malwares until it causes damage. Gartner [21] believes that eventually Signature-based techniques will be replaced with more robust approaches, because today the signature-based anti-malwares have marginal value as 2nd generation malwares can easily escape detection.

### 3.2 Heuristics based detection

In this method there are two approaches for the detection of malwares. Firstly in static approach suspicious program are disassembled to find a matching of the known malware pattern, if any. If the analysis result crosses the preset threshold then the program is marked as infected [22]. Secondly in dynamic approach, code emulation techniques are used by simulating the processor and operating system to detect suspicious operations (an attempt to open other executable files with the intention of modifying its content, changing the Master Boot Record, concealing themselves from the operating system, etc.) on a virtual machine.

The Heuristics method is a promising technique for the detection of unknown malware, in particular to detect encrypted malwares [20]. However, it requires entire virtual environment to be installed. Also it is prone to false alarm [23], which may make the system more vulnerable by taking the real malware as another false alarm. Researchers augment the results of detection techniques and combine it with another detection technique to reduce the false alarm [24].

### 3.3 Machine Learning

In recent years, malware detection with machine learning techniques is gaining popularity. Tom Mitchell defines machine learning as the study of computer algorithms that improve through experiments [25]. Robert Moskovitch et. al. proposed detection of malwares based on monitoring the computer behavior (features). His evaluation results suggest that by using classification algorithm applied on only 20 features the mean detection accuracy exceeded 90% [26]. The advantage of machine learning techniques is that it will not only detect known malwares but also act as knowledge for the detection of new malware. The popular machine learning techniques among the researchers for the detection of 2nd generation malwares are Naive Bayes [27], Decision Tree [28], Data Mining [29], Neural Networks [28] and Hidden Markov Modes [15].

This technique may not replace the standard detection methods, but can act as an add-on feature. Generally, machine learning techniques are more computationally demanding then the standard anti-malware, hence it may not be suitable for end users. However, it can be implemented at enterprise gateway level to act as a central anti-malware engine to supplement anti-malwares. Although, infrastructure requirement is costly, but it can help in protecting valuable enterprises data from the security threat and can prevent immense financial damages.

### 3.4 Malware Normalization

The malwares generated from advanced toolkits such as UPX and Mitsfall are difficult to detect [30]. For the detection of such malwares, normalization techniques can be used to improve the detection rate of an existing anti-malware. In this technique, normalizer accepts the obfuscated version of malware and eliminates the obfuscation carried on the program and produce the normalized executable. After normalization the signature of the malware is extracted and compared with the signature of canonical form [31]. Christodorescu et. al. designed a malware normalizer that handles three common obfuscations viz. code reordering, packing, and junk insertion [32]. Later on Armor et. al., [33] proposed a generalized malware normalizer which can store obfuscation methods in the form of automata structures and use them for normalizing the metamorphic mal- wares. Recently a general malware normalizer has been proposed that can store lots of obfuscation methods in the form of automata structures for normalizing metamorphic malwares, which has a detection rate up to 81% [33].

## 4. CONCLUSION

The malware creators are ahead of the anti-malware developer. The reason is the availability of good softwares to create variants of malwares [8]. So far for the detection of malwares; Signature matching [22], Heuristic approach [20], Machine learning [28] and Normalization methods [33] are used. There is no method available to detect zero day attack malwares with 100% accuracy [27]. Also, not much has been done to detect malwares of LINUX OS and PDAs, which are under the radar of malware attackers. It has been reported that there exists malwares which cannot be detected by any anti-malware [24]. Moreover it is impossible to develop a generic algorithm to detect all possible malwares [34]. Hence regular study is required to combat the threat/attacks from the new malwares. Therefore from time to time both academic community and software companies proposed methodologies and offer products to fight against malwares. Currently research groups are focusing on the detection of metamorphic malwares generated by NGVCK [35], Virus Construction Set and Genvir [8]. To detect the metamorphic malwares, normalization techniques can be exploited [33].

Malwares in PDAs are increasing at an unprecedented rate and it is mainly due to the ease of generating malware variants [7]. The recent attacks on PDAs show that there is an urgent need to develop robust anti-malwares, in particular for defense against zero-day attack [37]. The detection of malwares in PDAs (Android) is done primarily by permission leakage [38]. Min Zheng et. al., 2013 proposed a signature based analytic system to automatically collect, manage, analyze and extract android malware known as Droid Analytics. They used 150,368 Android applications and successfully determined 2,497 Android malwares from 102 different families, with 342 of being zero-day malware samples from six different families. However, there are still open questions viz. how to detect new malware variants in PDAs which are always hidden in the many different third-party markets [39]. Also it is important to find out how one can identify repackaged applications from the vast ocean of applications and malwares. In order to overcome the above issue cloud computation may be one of the solutions.

In addition, efficient malware detection plays an important role for the end users. Traditionally, malware detection techniques use large database installed in the system. Hence require to develop efficient methods for scanning the malwares. Recent NIST approved hash function called Secure Hash Algorithm - 3 (Keecak) may be used for efficient scanning [36]. Also one can use the power of general-purpose graphics processing unit for efficient scanning of malwares.

## 5. ACKNOWLEDGEMENTS
We are thankful to Prof. Bharat Deshpande and Prof. Neena Goveas for the useful discussions and valuable suggestions.